\documentclass[preprint,12pt]{elsarticle}
\usepackage{graphicx}
\usepackage{amssymb}
\usepackage{epsfig}
\usepackage{subfigure}

\journal{Physica A}

\begin{document}

\begin{frontmatter}

\title{A comparative study of the dynamic critical behavior of the four-state Potts like models}
\author[UFOP]{E. Arashiro}\ead{arashiro@iceb.ufop.br}
\author[UFG]{H. A. Fernandes}
\author[USP]{J. R. Drugowich de Fel\'{\i}cio}
\address[UFOP]{Universidade Federal de Ouro Preto, Departamento de F\'{\i}sica \\Campus Universit\'{a}rio Morro
do Cruzeiro, Ouro Preto, MG, Brazil, 35400-000}
\address[UFG]{Universidade Federal de Goi\'{a}s, Campus Jata\'{\i} \\BR 364, Km 192, n. 3800, C.P. 03, Setor
Parque Industrial, Jata\'{\i}, GO, Brazil, 78000-000}
\address[USP]{Universidade de S\~{a}o Paulo, Faculdade de Filosofia, Ci\^{e}ncias e Letras de Ribeir\~{a}o Preto \\
Avenida Bandeirantes, n. 3900, Ribeir\~{a}o Preto, SP, Brazil, 14040-901}

\begin{abstract}
We investigate the short-time critical dynamics of the Baxter-Wu
(BW) and $n=3$ Turban (3TU) models to estimate their global
persistence exponent $\theta _{g}$. We conclude that this new
dynamical exponent can be useful in detecting differences between
the critical behavior of these models which are very difficult to
obtain in usual simulations. In addition, we estimate again the
dynamical exponents of the four-state Potts (FSP) model in order to
compare them with results previously obtained for the BW and 3TU
models and to decide between two sets of estimates presented in the
current literature. We also revisit the short-time dynamics of the
3TU model in order to check if, as already found for the FSP model,
the anomalous dimension of the initial magnetization $x_{0}$ could
be equal to zero.
\end{abstract}

\end{frontmatter}

\section{INTRODUCTION}

Since the works by Janssen, Schaub and Schmittmann
\cite{Janssen1989}, and Huse \cite{Huse1989}, the critical
properties of statistical systems have been a subject of
considerable interest in nonequilibrium physics
\cite{Li1995,Tome1998a,Zheng1998,Grandi2004,Arashiro2006,Lei2007,
Dasilva2002b,Asad2007,Kornyei2008,Nobre2008,Nam2008}. By using
renormalization group methods and numerical calculations,
respectively, they showed that there is universality and scaling
behavior even at the early stage of the time evolution after
quenching from high temperatures to the critical one.

The dynamic scaling relation obtained by Janssen \textit{et al.}
\cite{Janssen1989} for the \textit{k}th moment of the magnetization,
extended to systems of finite size \cite{Li1995}, is written as
\begin{equation}
M^{(k)}(t,\tau ,L,m_{0})=b^{-k\beta /\nu }M^{(k)}(b^{-z}t,b^{1/\nu }\tau
,b^{-1}L,b^{x_{0}}m_{0}),  \label{Eq:scaling}
\end{equation}
where $t$ is the time evolution, $b$ is an arbitrary spatial scaling factor,
$\tau =(T-T_{c})/T_{c}$ is the reduced temperature and $L$ is the linear
size of the lattice. The exponents $\beta $ and $\nu $ are as usual the
equilibrium critical exponents associated respectively with the order
parameter and the correlation length, $z$ is the dynamical exponent
characterizing time correlations in equilibrium, and $x_{0}$ represents the
anomalous dimension of the initial magnetization $m_{0}$, introduced to
describe the dependence of the scaling behavior on the initial conditions.

Besides to avoid the well-known problem of the \textquotedblleft critical
slowing down\textquotedblright , characteristic of the equilibrium, and to
provide an alternative way to obtain the familiar set of static critical
exponents and the dynamic critical exponent $z$, this kind of investigation
reveals a new universal regime and an unsuspected new dynamic critical
exponent $\theta $ which can be found by following the above scaling law for
the order parameter at the critical temperature ($\tau =0$)
\begin{equation}
M(t)\thicksim m_{0}t^{\theta }.  \label{Eq:theta}
\end{equation}

This new index, independent of the previously known exponents
characterizes the so-called \textquotedblleft critical initial
slip\textquotedblright, the anomalous behavior of the order
parameter when a system is quenched to the critical temperature
$T_{c}$. This exponent is related to $x_{0}$ as
\begin{equation}
\theta =\frac{x_{0}-\beta /\nu }{z}.  \label{Eq:x0}
\end{equation}

Some years later, Majumdar \textit{et al.} \cite{Majumdar1996} have
shown that another dynamic critical exponent can be obtained in the
study of systems far from equilibrium. By studying the behavior of
the global persistence probability $P(t)$ that the order parameter
has not changed its sign up to time $t$, they have
shown that $P(t)$ should behave, at the critical temperature, as
\begin{equation}
P(t)\thicksim t^{-\theta _{g}},  \label{Eq:thetag}
\end{equation}
where $\theta _{g}$ is the global persistence exponent. They also
argued that, if the time evolution of the order
parameter would be a Markovian process, then the exponent $\theta
_{g}$ should obey the equation \cite{Majumdar1996}
\begin{equation}
\theta _{g}=\alpha _{g}=-\theta +\frac{d}{2z}-\frac{\beta }{\nu z}.
\label{Eq:alpha}
\end{equation}
However, as shown in several works \cite{Majumdar1996,Majumdar2003,
Schulke1997,Oerding1997,Ren2003,Albano2001,
Saharay2003,Hinrichsen1998,Sen2004,Zheng2002,Fernandes2006a,Fernandes2006b,
Fernandes2006c} the exponent $\theta _{g}$ is an independent critical index
closely related to the non-Markovian characteristic of the process.

In this work, we perform short-time Monte Carlo simulations to
investigate the scaling behavior of the global persistence
probability $P(t)$ for the BW \cite{Wood1972,Baxter1973}, 3TU
\cite{Turban1982,Turban1983} and FSP \cite{Potts1952,Wu1992} models
in two dimensions ($d=2$), that exhibit the same set of leading
static critical exponents. We also calculate the exponent $x_{0}$ of
these models but only after reobtaining more precise estimates for
the dynamical indices $\theta $ and $z$ related to the FSP and 3TU
models. The aim of this paper is to show that is also possible to
detect different behavior between those models by doing short-time
Monte Carlo simulations.

The paper is organized as follows. In the next section we present
the models. In Section \ref{section3} we show the short-time scaling
relations and present our results. Finally, in Section
\ref{section4} we present our conclusions.

\section{THE MODELS} \label{section2}

The $q-$state Potts model which is a simple extension of the Ising model,
has a rich phase diagram \cite{Wu1992} with first order phase transitions
when $q>4$ and second order phase transitions when $q\leq 4$. Its
Hamiltonian is given by
\begin{equation}
-\beta \mathcal{H}=K\sum_{\langle i,j\rangle }\delta _{\sigma _{i}\sigma
_{j}},
\end{equation}
where $\beta =1/k_{B}T$ and $k_{B}$ is the Boltzmann constant, $\langle
i,j\rangle $ represents nearest-neighbor pairs of lattice sites, $K$ is the
dimensionless ferromagnetic coupling constant and $\sigma _{i}$ is the spin
variable which takes the values $\sigma _{i}=0,\cdots ,q-1$ on the lattice
site $i$. It is well known that the critical coupling of this model is given
by \cite{Wu1992}
\begin{equation}
K_{c}=\mbox{log}(1+\sqrt{q}),
\end{equation}
and its order parameter is defined as
\begin{equation}
M=\frac{1}{L^{d}(q-1)}\left\langle \sum_{i}(q\delta _{\sigma
_{i}(t),1}-1)\right\rangle  \label{Eq:FSPM}
\end{equation}
where $L$ is the linear size of the lattice and $d$ is the dimension of the
system. The case $q=4$ (FSP model) in two dimensions is known to exhibit
slow convergence when investigated by finite-size techniques motivated by
the presence of a marginal operator (scaling dimension = d = 2).

The BW model is defined by the Hamiltonian
\begin{equation}
-\beta \mathcal{H}=K\sum_{\langle i,j,k\rangle }\sigma _{i}\sigma _{j}\sigma
_{k},  \label{Eq:BW}
\end{equation}
where $\sigma _{i}=\pm 1$ is an Ising spin
variable located at each site of
the triangular lattice and the sum extends over all elementary
triangles.

The Hamiltonian of the 3TU model is given by
\begin{equation}
-\beta \mathcal{H}=\sum_{\langle i,j\rangle }\left\{ K_{h}\sigma
_{i,j}\sigma _{i+1,j}\sigma _{i+2,j}+K_{v}\sigma _{i,j}\sigma
_{i,j+1}\right\} ,  \label{Eq:TU}
\end{equation}
where the sum is over all sites of a square lattice, $K_{h}$ and
$K_{v}$ are the coupling constant in the horizontal (with three-spin
interactions) and vertical (with two-spin interactions) directions,
respectively, and $\sigma _{i,j}=\pm 1$ is an Ising
spin variable located at each site of the lattice.

Both BW and 3TU (for the isotropic case, $K_{h}=K_{v}=K$) models
undergo a continuous phase transition at the critical temperature
$K_{c}=0.5\ln (1+ \sqrt{2})$ which is the same critical temperature
of the Ising model on a square lattice. The order parameter of these
models is defined as
\begin{equation}
M=\frac{1}{L^{d}}\left\langle \sum_{i}\sigma _{i}\right\rangle .
\label{Eq:BWTUM}
\end{equation}

The BW and 3TU models present semi-global up-down spin reversal symmetry
\cite{Alcaraz1987}, i.e., their Hamiltonians are invariant under reversal of
all the spins belonging to two of three sublattices into which the original
lattice can be decomposed.

The ground state of these three models is fourfold degenerated,
being that the possible spin configurations of the BW and 3TU models
consist of repetitions of the patterns $\{+,+,+\}$, $\{+,-,-\}$,
$\{-,+,-\}$ or $\{-,-,+\}$. The main difference between the three
models is that the BW model is defined on a triangular lattice
whereas in the FSP and 3TU models the spins are
located on a square one.

From the degeneracy and symmetry considerations, it was conjectured
that these three models would belong to the same universality class,
with critical exponents given by \cite{Baxter1982}
\begin{equation}
\beta =\frac{1}{12},\hspace{0.3cm}\nu =\alpha
=\frac{2}{3},\hspace{0.3cm} \mbox{and}\hspace{0.3cm}\eta
=\frac{1}{4}.  \label{Eq:static}
\end{equation}

However, when these models are deeply studied, differences among
sub-dominant exponents appear. These exponents are supposed to be
associated to different behavior exhibited by those models when
studied by finite-size scaling techniques. This fact was first
pointed out by Alcaraz and Xavier \cite{Alcaraz1997} in a
finite-size scaling study of the FSP and BW models using a conformal
invariance approach.

As will be shown in this paper, it is possible to observe remarkable
differences between those models by investigating
the non-equilibrium evolution of a dynamical quantity introduced by
Majumdar et al. \cite{Majumdar1996}, the global
persistence probability. \ This result corroborates previous
simulations which pointed out different behavior for the BW model
when compared to the FSP model
\cite{Santos2001,Arashiro2003,Chatelain2004}.

\section{RESULTS AND DISCUSSIONS}

\label{section3}

In our Monte Carlo simulations, we consider two-dimensional lattices with
periodic boundary conditions. The dynamical evolution of the spins is local
and updated by the heat-bath algorithm at the critical coupling $K_{c}$. In
order to check finite-size effects, we consider three different lattice
sizes ($L=120$, $180$ and $240$) the exponents being obtained from five
independent bins of 20000 samples each one.

\subsection{Global persistence exponent $\protect\theta_g$}

The global persistence probability $P(t)$ can be defined as
\begin{equation}
P(t)=1-\sum_{t^{\prime}=1}^{t}\rho(t^{\prime})  \label{eq:probability}
\end{equation}
where $\rho(t^{\prime})$ is the fraction of the samples that have changed
their state for the first time at the instant $t^{\prime}$. The dynamical
exponent $\theta_g$ that governs the behavior of $P(t)$ at criticality is
obtained through the power law behavior given by
\begin{equation}
P(t) \thicksim t^{-\theta_g}.
\end{equation}

In order to obtain the exponent $\theta_g$, the initial configuration of the
system should be carefully prepared with a precise and small value of $m_0$.
After estimating $\theta_g$ for a number of $m_0$ values, its final value is
obtained from the limit $m_0 \rightarrow 0$. In this work, we used $4\cdot
10^{-4} < m_0 \leq 5\cdot 10^{-3}$.

As we are considering three models and two different order
parameters, it is worth to explain how to obtain
$m_{0}$ in each case. At the beginning, each site on the lattice of
the FSP model is occupied by a spin variable which takes the values
$\sigma =0$, 1, 2 or 3 and, for the 3TU and BW
models, the sites are occupied by spin variables which take the
values $\pm 1$. For each model, the values of spin
variables are chosen with equal probability.
Afterward, the magnetization of the models is
measured by using the Eqs. (\ref{Eq:FSPM}) for the FSP model or
(\ref{Eq:BWTUM}) for the 3TU and BW models. In
order to obtain a null value of the initial magnetization, some
sites of the lattices are randomly chosen and its signs (or values)
are changed. Finally, the desired value for the initial
magnetization of each model is obtained by changing the sign (or the
value) of $\delta $ sites on the lattice. When using the Eq.
(\ref{Eq:FSPM}), the initial magnetization is given by
\begin{equation}
m_{0}=\frac{4\delta }{3L^{2}}  \label{m_01}
\end{equation}
and a value of $m_{0}$ is obtained choosing $\delta $ sites occupied
by $\sigma =0$, 2 or 3 and substituting them by $\sigma =1$. For the
Eq. (\ref{Eq:BWTUM}), $m_{0}$ is simply given by
\begin{equation}
m_{0}=\frac{\delta }{L^{2}}  \label{m_02}
\end{equation}
and to obtain a value of $m_{0}$, we choose randomly $\delta /2$ sites
occupied by $\sigma =-1$ and substitute them by $\sigma =1$.

\begin{figure}[ht]
\begin{center}
\subfigure{\epsfig{file=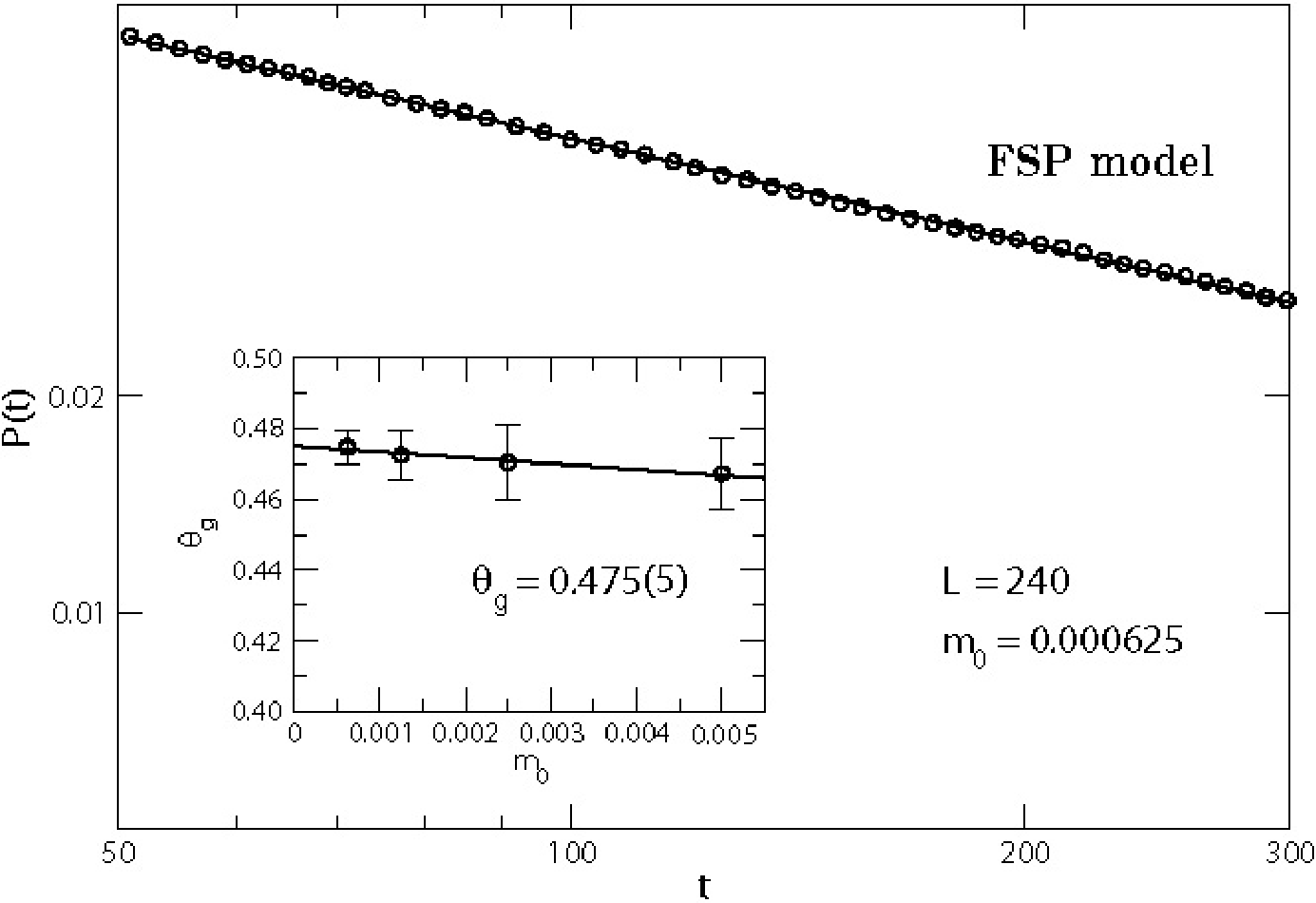,height=6cm,width=8.5cm}} \newline
\subfigure{\epsfig{file=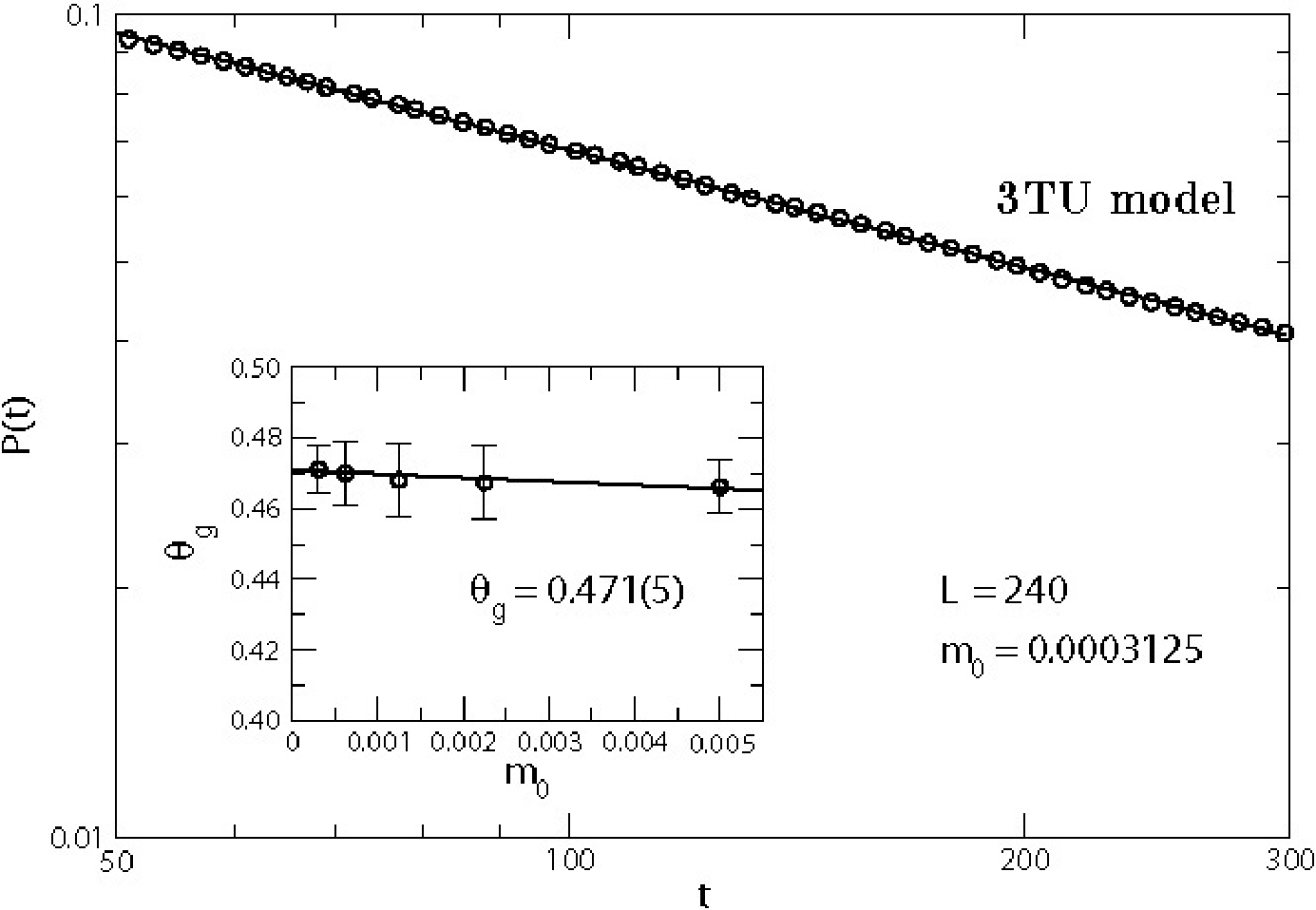,height=6cm,width=8.5cm}} \newline
\subfigure{\epsfig{file=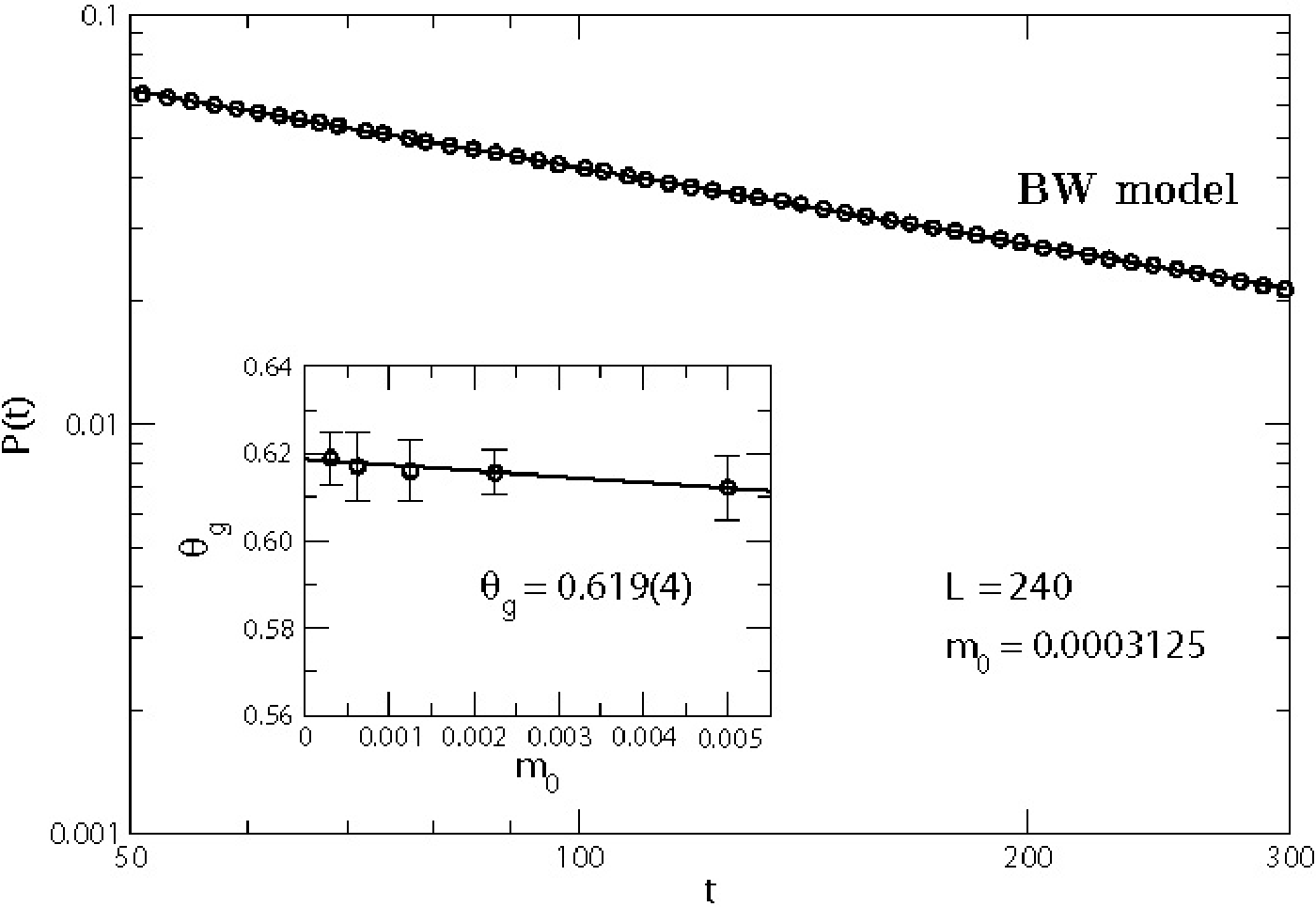,height=6cm,width=8.5cm}}
\end{center}
\caption{The time evolution of the global persistence probability
$P(t)$ for $L=240$ for the FSP (on top), 3TU (on
middle), and BW (on bottom) models. The error bars calculated over 5
sets of $20000$ samples are smaller than the symbols. The inset in
each figure shows the exponent $\protect\theta_g$ for different
initial magnetizations, as well as its extrapolated value.}
\label{Fig:theta_g}
\end{figure}

In Fig. \ref{Fig:theta_g} we show the behavior of the global
persistence probability for $L=240$ and a small value of $m_{0}$ for
the FSP(on top), 3TU (on middle), and BW (on
bottom) models, in double-log scales. The error bars, calculated
over five sets of $20000$ samples are smaller than the symbols.

The insets in Fig. \ref{Fig:theta_g}
display the estimates of $\theta _{g}$ for
different values of $m_{0}$ and the limiting procedure
$m_{0}\rightarrow 0$ for the models.

In Table \ref{Tb:persistence}, we show the extrapolated values of
$\theta _{g}$ for $L=120$, $180$ and $240$ for the BW, 3TU and FSP
models. Finite-size effects are less than statistical errors.

The discrepancy among the results obtained for the persistence
exponent for the BW model by one side and for the FSP and 3TU models
is noteworthy. Discrepancies between the BW and FSP models in
dynamical simulations were previously observed by Arashiro and
Drugowich de Fel\'{\i}cio \cite{Arashiro2003} for the dynamical
exponent $\theta $ and by Chatelain \cite{Chatelain2004} for the
exponent $\lambda /z=d/z-\theta $ and for the asymptotic value of
the fluctuation-dissipation ratio $X_{\infty }$. However, from
numerical calculations made on finite lattices (at the equilibrium)
it is well known that BW, FSP and 3TU models show different
corrections to finite-size scaling. Whereas estimates for the BW
model exhibit good convergence with the system size
\cite{Alcaraz1997}, comparable to that of the two-dimensional Ising
model, FSP and 3TU models offer serious barriers to whom wish to
find their exponents from finite-size techniques
\cite{Igloi1983,Igloi1987}.

\subsection{Dynamic critical exponents $\protect\theta$ and $z$}

As conjectured by Janssen \textit{et al.} \cite{Janssen1989} on the
basis of renormalization group techniques and by Huse
\cite{Huse1989} through numerical calculations, in the short-time
regime, the order parameter obeys a power law as shown in the Eq.
(\ref{Eq:theta}). Formerly, a positive value was always associated
to this exponent
\cite{Zheng1998,Jaster1999,Schulke1995,Tome1998,Fernandes2005} and
the phenomenon was known as critical initial slip. However, as shown
in some papers, there are models in which the exponent $\theta $ can
have a negative value, for instance, the tricritical Ising model,
\cite{Janssen1994,Dasilva2002}, FSP \cite{Dasilva2004}, 3TU
\cite{Simoes2001}, and BW \cite{Arashiro2003,Malakis2005} models.

Although the estimates of the exponents $\theta $ and $z$ for the BW
model are known with good precision, the results for the 3TU model
exhibit large error bars. In addition, estimates obtained for
$\theta $ in previous papers \cite{Fernandes2006b,Dasilva2004} show
considerable differences between the two techniques employed to
study the FSP model. So, in order to obtain more precise estimates
for the exponent $x_{0}$, we decided to reobtain the exponents
$\theta $ and $z$ for both FSP and 3TU models by using the time
correlation of the magnetization \cite{Tome1998}
\begin{equation}
C(t)=\langle M(0)M(t)\rangle \thicksim t^{\theta }  \label{Eq:TC}
\end{equation}
and the function $F_{2}(t)$ proposed by da Silva \textit{et al.}
\cite{Dasilva2002b}
\begin{equation}
F_{2}(t)=\frac{\langle M^{2}(t)\rangle _{m_{0}=0}}{\langle M(t)\rangle
_{m_{0}=1}^{2}}\thicksim t^{d/{z}}.  \label{Eq:f2}
\end{equation}
In Eq. (\ref{Eq:TC}), the average is taken over a set of random
initial configurations. Initially, this approach had shown to be
valid only for models which exhibit up-down symmetry
\cite{Tome1998}. Nevertheless, it has been later found that this
approach is more general and can include models with other
symmetries \cite{Tome2003}. This method has several advantages when
compared to other approaches, for instance, the exponent $\theta $
can be directly calculated without the need of careful preparation
of the initial states nor of the limiting procedure [see Eq.
(\ref{Eq:theta})], the only requirement being that $\langle
M(0)\rangle =0$.

\begin{figure}[ht]
\begin{center}
\subfigure{{\epsfig{file=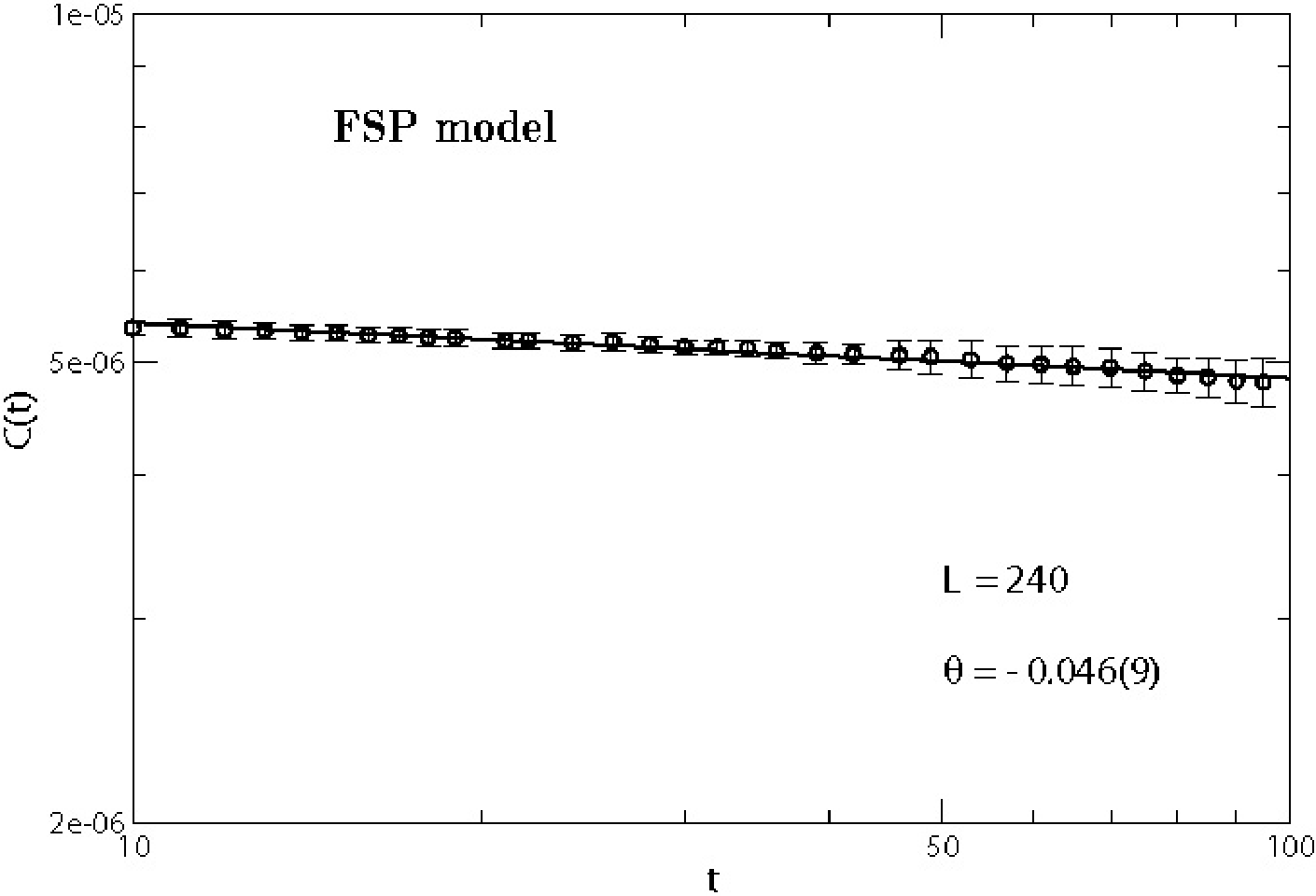,height=6cm,width=8.5cm}}} \\[0pt]
\subfigure{{\epsfig{file=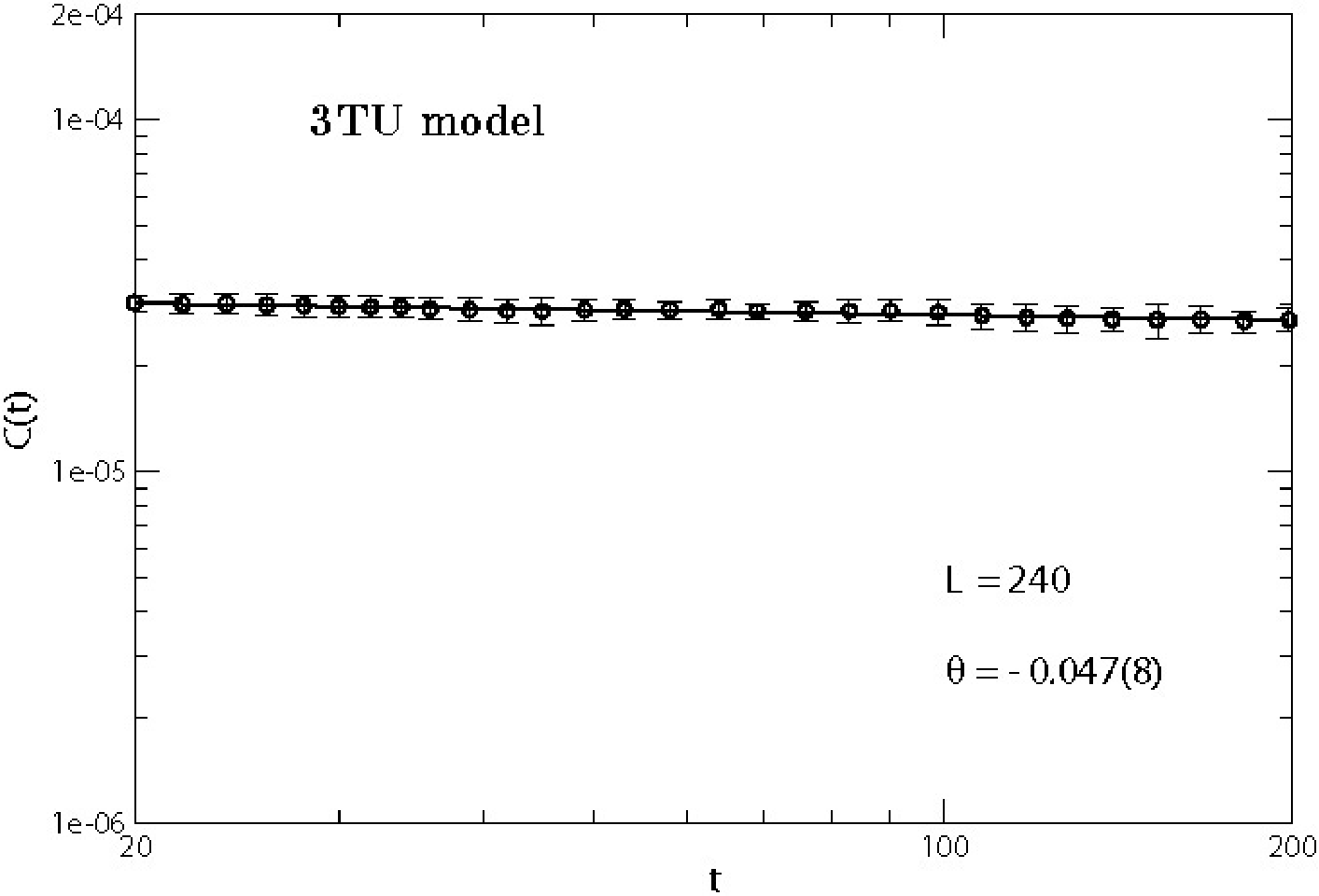,height=6cm,width=8.5cm}}}
\end{center}
\caption{The time correlation of the order parameter on log
scales for the FSP (on top) and
3TU (on bottom) models. Error bars were calculated
over 5 sets of $20000$ samples.} \label{Fig:theta}
\end{figure}

In Fig. \ref{Fig:theta} we show the evolution of the time
correlation $C(t)$ in double-log scale for the FSP (on top) and
3TU (on bottom) models, respectively, for $L=240$.

The slope of these curves is shown in Table \ref{Tb:theta}, as well
as the estimates for $L=120$ , $180$ and $240$.

\begin{figure}[ht]
\begin{center}
\subfigure{{\epsfig{file=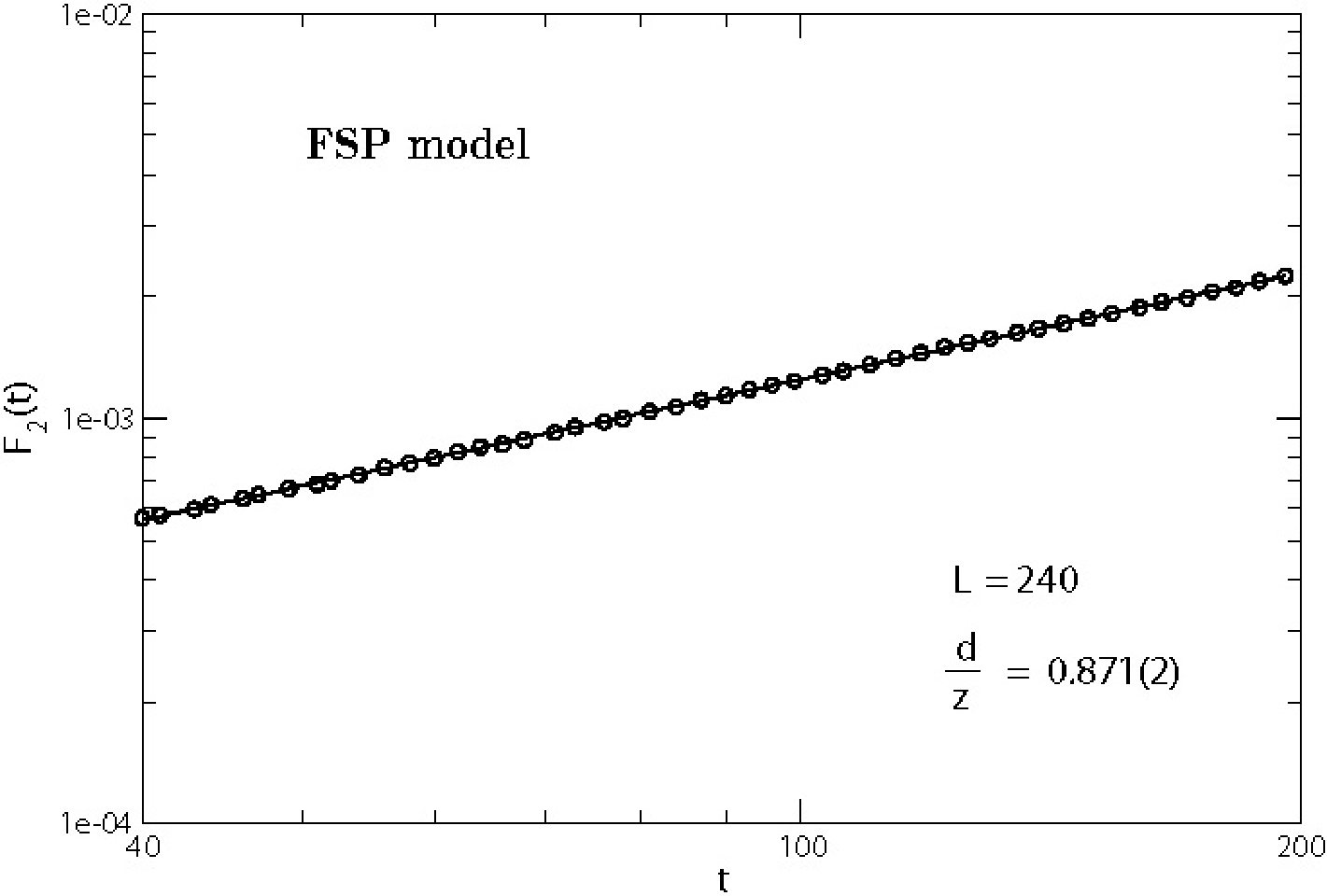,height=6cm,width=8.5cm}}} \\[0pt]
\subfigure{{\epsfig{file=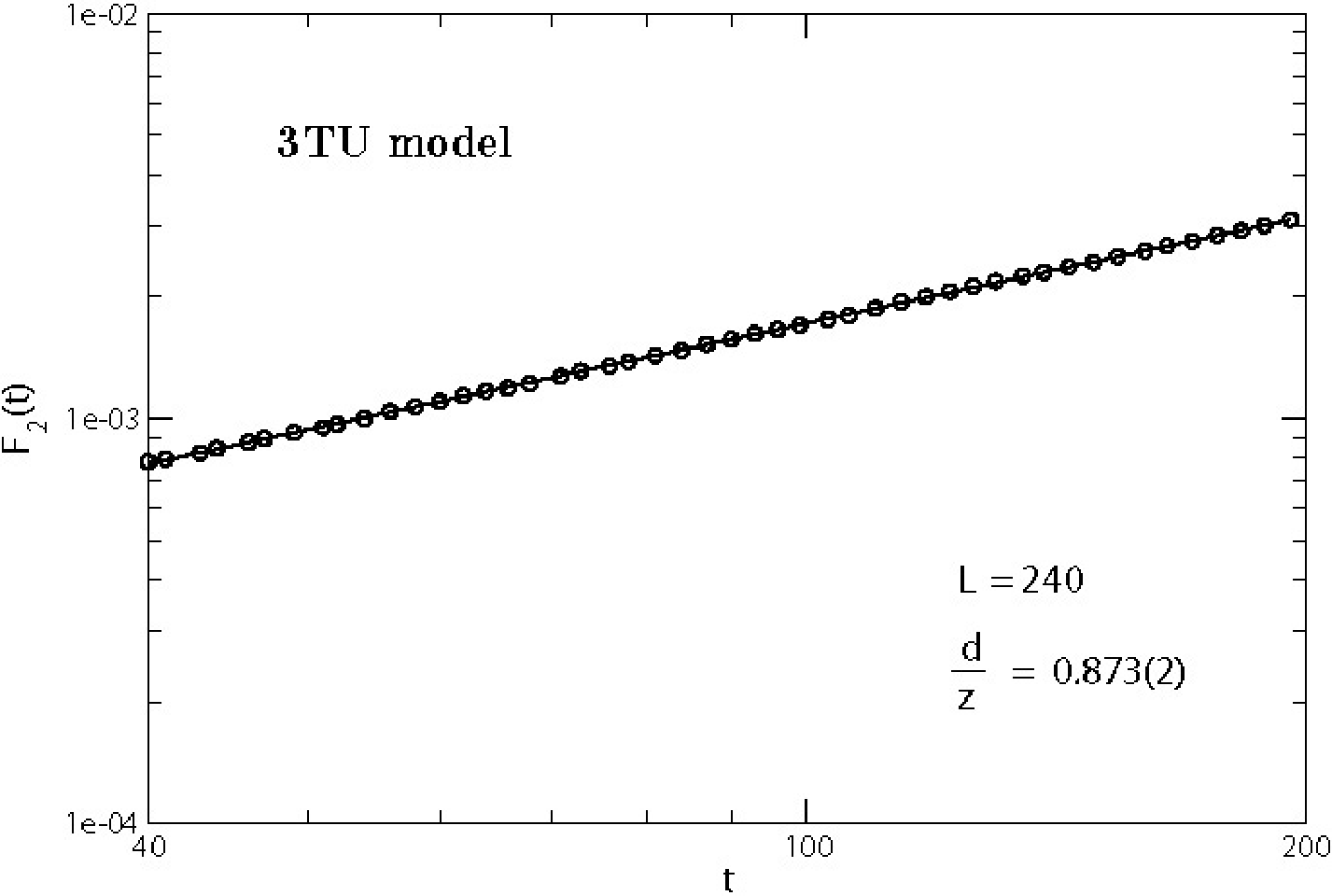,height=6cm,width=8.5cm}}}
\end{center}
\caption{The time evolution of $F_2(t)$ for $L=240$ for the FSP (on
top) and 3TU (on bottom) models. Each point
represents an average over 5 sets of $20000$ samples and the error
bars are obtained of them.} \label{Fig:f2}
\end{figure}

On the other hand, the dynamical exponent $z$ was obtained by
combining results from samples submitted to different initial
conditions (see Eq. (\ref{Eq:f2}), where $d=2$ is the dimension of
the system). This approach has proved to be very efficient in
estimating the exponent $z$ for several models
\cite{Dasilva2002b,Dasilva2002,Arashiro2003,Fernandes2005}. The time
evolution of $F_2$ is shown on log scales in Fig. \ref{Fig:f2} for
$L=240$ for the FSP (on top) and 3TU (on bottom)
models.

Taking into account the values of the ratio $d/z$, estimated from
the slope of these curves, the exponent $z$ can be easily found. Our
estimates for this exponent for the FSP and 3TU models are shown in
Table \ref{Tb:z} for $L=120$, $180$, and $240$.

As shown in Table \ref{Tb:z}, the estimates for the dynamical
exponent $z$ of the FSP and 3TU models are in complete agreement
with our results for the BW model \cite{Arashiro2003}
($z=2.294(6)$). However, the values we found for the dynamical
exponent $\theta$ of the FSP and 3TU models (Table \ref{Tb:theta})
are completely different from the previously estimated exponent for
the BW model \cite{Arashiro2003}
\begin{equation}
\theta =-0.186(2).  \label{Eq:thetazBW1}
\end{equation}

\subsection{The exponent $\protect\alpha _{g}$ and the anomalous dimension $x_{0}$}

Using the results of the Tables \ref{Tb:theta} and \ref{Tb:z} for
$L=240$ (FSP and 3TU models), the results of the
Eq. (\ref{Eq:thetazBW1}) and the values of $\beta $
and $\nu $ of Eq. (\ref{Eq:static}) we estimate the exponent $\alpha
_{g}$ through Eq. (\ref{Eq:alpha}) for the studied models (see Table
\ref{Tb:alpha}). The difference between our estimate for $\theta
_{g}$ (See Table \ref{Tb:persistence}) and the value obtained from
Eq. (5) shows the non-Markovian aspect for the BW, 3TU and FSP
models. Thus, the global persistence exponent in these cases is also
independent of other critical exponents.

We remark that using the estimates of Hadjiagapiou \textit{et al.}
\cite{Malakis2005} for the dynamical exponents of the BW model
$z=1.994(24)$ and $\theta=-0.185(2)$ we obtain $\alpha
_{g}=0.624(3)$ approximately equal to $\theta _{g}$ (from Table
\ref{Tb:persistence}) which means that the relaxation would be
Markovian. As we know, the models studied until now
\cite{Majumdar1996,Majumdar2003,
Schulke1997,Oerding1997,Ren2003,Albano2001,
Saharay2003,Hinrichsen1998,Sen2004,Zheng2002,Fernandes2006a,Fernandes2006b,
Fernandes2006c} exhibit different values for those exponents and
this would be the first case where Eq. (\ref{Eq:alpha}) would be
valid.

Finally, we calculate the value of the anomalous dimension $x_{0}$
of the order parameter for the FSP, 3TU, and BW models. This
exponent, which is introduced to describe the dependence on the
scaling behavior of the initial conditions, is related to the
exponents $\theta $, $z$, and $\beta /\nu $ by\ Eq. (\ref{Eq:x0}).
So,
\begin{equation}
x_{0}=\theta z+\beta /\nu .
\end{equation}
In Table \ref{Tb:x0}, we show the estimates of $x_{0}$ by using the
values of the Tables \ref{Tb:theta} and \ref{Tb:z} for $L=240$ (FSP
and 3TU models), the estimates for $\theta $
\cite{Arashiro2003,Malakis2005} and the conjectured values of $\beta
$ and $\nu $ given by Eq. (\ref{Eq:static}).

As we can see in Table \ref{Tb:x0}, our results indicate that far
from equilibrium the critical behavior of the FSP and 3TU models is
very similar but different from the BW one. In addition, our
estimates do not exclude a null value for the anomalous dimension of
the magnetization ($x_{0}$) in those cases (FSP and 3TU models)
which in static critical phenomena theory is known to be associated
to marginal operators \cite{Wegner1976} and in finite-size scaling
calculations to logarithmic corrections \cite{Kenna1986,Reis1996}.

\section{CONCLUSIONS}

\label{section4}

We estimated the dynamical exponent $\theta _{g}$ for the BW, 3TU,
and FSP models using the time evolution of the global persistence
probability that the magnetization has not changed its signal up to
time $t$. The value of $\theta _{g}$ found for the BW is completely
different from that found for the 3TU and FSP models. On the other
hand our results for the FSP and 3TU models are in good agreement
each other. As previously found for the exponent $\theta $ of the
initial magnetization, the persistence exponent of Majumdar et al.
\cite{Majumdar1996} is also able to detect differences between\ the
BW and the 3TU and FSP models. We stress that those differences are
very difficult to obtain in static critical phenomena study. We also
reobtained the dynamical exponent $\theta $ for the 3TU and FSP
models. In the case of the 3TU model we found much more precise
values than previously found by Sim\~{o}es and Drugowich de Felicio
\cite{Simoes2001} in consequence of better statistics whereas
present results for the FSP model only confirm estimates recently
found by Fernandes et al. \cite{Fernandes2006b} using a different
order parameter introduced by Vanderzande \cite{Vanderzande}.
Finally, with new and more precise estimates for the exponent
$\theta $ we could recalculate the anomalous dimension $x_{0}$ of
the initial magnetization for the three models and the conclusion is
that the value zero (related to a marginal operator in the language
of the renormalization group) can not be discarded for FSP and 3TU
models but is completely unlikely for the BW model.

\section*{Acknowledgments}

This work was supported by the Brazilian agencies CAPES, CNPq and
FAPESP. One of us (H.A.F.) would like to thank to the Universidade
Federal de Goi\'{a}s (PRPPG and FUNAPE).

\newpage


\newpage
\begin{table}[!htb]
\caption{The global persistence exponent $\protect\theta_g$ from the
power law behavior for the FSP, 3TU, and BW
models.} \label{Tb:persistence}\centering
\begin{tabular}{cccc}
\hline\hline
~~Models & ~~~ $L=120$ & ~~~ $L=180$ & ~~~ $L=240$~~ \\ \hline
~~FSP & ~~~ 0.469(4) & ~~~ 0.472(6) & ~~~ 0.475(5)~~ \\
~~3TU & ~~~ 0.469(6) & ~~~ 0.470(5) & ~~~ 0.471(5)~~ \\
~~BW & ~~~ 0.620(5) & ~~~ 0.618(5) & ~~~ 0.619(4)~~ \\ \hline\hline
\end{tabular}
\end{table}
\begin{table}[tbh]
\caption{The exponent $\protect\theta $ for the FSP and
3TU models.} \label{Tb:theta}\centering
\begin{tabular}{ccc}
\hline\hline ~~~ $L$ & ~~~~~ FSP model & ~~~~~ 3TU
model ~~~ \\ \hline
~~~ 120 & ~~~~~ $-0.046(8)$ & ~~~~~ $-0.047(7)$ ~~~ \\
~~~ 180 & ~~~~~ $-0.047(8)$ & ~~~~~ $-0.046(7)$ ~~~ \\
~~~ 240 & ~~~~~ $-0.046(9)$ & ~~~~~ $-0.047(8)$ ~~~ \\ \hline\hline
\end{tabular}
\end{table}
\begin{table}[!htb]
\caption{The exponent $z$ for the FSP and 3TU
models.} \label{Tb:z}\centering
\begin{tabular}{ccc}
\hline\hline ~~~ $L$ & ~~~~~ FSP model & ~~~~~ 3TU
model ~~~ \\ \hline
~~~ 120 & ~~~~~ $2.294(7)$ & ~~~~~ $2.293(5)$ ~~~ \\
~~~ 180 & ~~~~~ $2.294(5)$ & ~~~~~ $2.290(8)$ ~~~ \\
~~~ 240 & ~~~~~ $2.296(5)$ & ~~~~~ $2.292(4)$ ~~~ \\ \hline\hline
\end{tabular}
\end{table}
\begin{table}[tbh]
\caption{The exponent $\protect\alpha _{g}$ for the FSP,
3TU, BW models.} \label{Tb:alpha}
\centering
\begin{tabular}{cc}
\hline\hline ~~~ Models & ~~~~~ $\alpha_{g}$ ~~~ \\ \hline
~~~ FSP & ~~~~~ $0.427(10)$~~ \\
~~~ 3TU & ~~~~~ $0.429(9)$ ~~~ \\
~~~ BW \cite{Arashiro2003} & ~~~~~ $0.567(3)$ ~~~ \\ \hline\hline
\end{tabular}
\end{table}
\begin{table}[tbh]
\caption{The exponent $x_{0}$ for the FSP, 3TU, BW
models.} \label{Tb:x0}\centering
\begin{tabular}{cc}
\hline\hline
~~~ Models & ~~~~~ $x_0$ ~~~ \\ \hline
~~~ FSP & ~~~~~ $0.019(21)$ ~~~ \\
~~~ 3TU & ~~~~~ $0.017(18)$ ~~~ \\
~~~ BW \cite{Arashiro2003} & ~~~~~ $-0.302(6)$ ~~~ \\
~~~ BW \cite{Malakis2005} & ~~~~~ $-0.244(8)$ ~~~ \\ \hline\hline
\end{tabular}
\end{table}


\newpage
\begin{thebibliography}{00}
\bibitem{Janssen1989} H. K. Janssen, B. Schaub, and B. Schmittmann, Z. Phys.
B: Condens. Matter \textbf{73}, 539 (1989).

\bibitem{Huse1989} D. A. Huse, Phys. Rev. B \textbf{40}, 304 (1989).

\bibitem{Li1995} Z. B. Li, L. Schulke and B. Zheng, Phys. Rev. Lett. \textbf{74}, 3396 (1995).

\bibitem{Tome1998a} T. Tome and J. R. Drugowich de Fel\'{\i}cio, Mod. Phys.
Lett. B \textbf{12}, 873 (1998).

\bibitem{Zheng1998} B. Zheng, Int. J. Mod. Phys B \textbf{12}, 1419 (1998).

\bibitem{Grandi2004} B. C. S. Grandi and W. Figueiredo, Phys. Rev. E \textbf{70}, 056109 (2004).

\bibitem{Arashiro2006} E. Arashiro, J. R. Drugowich de Fel\'{\i}cio, and U.
H. E. Hansmann, Phys. Rev. E \textbf{73}, 040902 (2006); J. Chem.
Phys. \textbf{126}, 045107 (2007).

\bibitem{Lei2007} X. W. Lei and B. Zheng, Phys. Rev. E \textbf{75}, 040104
(2007).

\bibitem{Dasilva2002b} R. da Silva, N. A. Alves, and J. R. Drugowich de
Fel\'{\i}cio, Phys. Lett. A \textbf{298}, 325 (2002).

\bibitem{Asad2007} A. Asad, B. Zheng, J. Phys. A: Math. Theor., \textbf{40},
9957 (2007).

\bibitem{Kornyei2008} L. K\"{o}rnyei, M. Pleimling, and F. Igloi, Phys. Rev.
E, \textbf{77}, 011127 (2008).

\bibitem{Nobre2008} S. D. da Cunha, U. L. Fulco, L. R. da Silva, and F. D. Nobre,
Eur. Phys. J. B, \textbf{63}, 93 (2008).

\bibitem{Nam2008} K. Nam, B. Kim, and S. J. Lee, Phys. Rev. E, \textbf{77}, 056104 (2008).

\bibitem{Majumdar1996} S. N. Majumdar, A. J. Bray, S. J. Cornell and C.
Sire, Phys. Rev. Lett. \textbf{77}, 3704 (1996).

\bibitem{Majumdar2003} S. N. Majumdar and A. J. Bray, Phys. Rev. Lett.
\textbf{91}, 030602 (2003).

\bibitem{Schulke1997} L. Schulke and B. Zheng, Phys. Lett. A \textbf{233},
93 (1997).

\bibitem{Oerding1997} K. Oerding, S. J. Cornell, and A. J. Bray, Phys. Rev.
E \textbf{56}, R25 (1997).

\bibitem{Ren2003} F. Ren and B. Zheng, Phys. Lett. A \textbf{313}, 312
(2003).

\bibitem{Albano2001} E. V. Albano and M. A. Mu\~{n}oz, Phys. Rev. E \textbf{63}, 031104 (2001).

\bibitem{Saharay2003} M. Saharay and P. Sen, Phys. A \textbf{318}, 243
(2003).

\bibitem{Hinrichsen1998} H. Hinrichsen and H. M. Koduvely, Eur. Phys. J. B
\textbf{5}, 257 (1998).

\bibitem{Sen2004} P. Sen and S. Dasgupta, J. Phys. A: Math. Gen. \textbf{37}, 11949 (2004).

\bibitem{Zheng2002} B. Zheng, Mod. Phys. Lett. B \textbf{16}, 775 (2002).

\bibitem{Fernandes2006a} H.A. Fernandes and J.R. Drugowich de Fel\'{\i}cio,
Phys. Rev. E \textbf{73}, 57101 (2006).

\bibitem{Fernandes2006b} H. A. Fernandes, E. Arashiro, J. R. Drugowich de Fel\'{\i}cio,
and A. A. Caparica, Physica A \textbf{366}, 255 (2006).

\bibitem{Fernandes2006c} H. A. Fernandes, Roberto da Silva, and J. R.
Drugowich de Fel\'{\i}cio, J. Stat. Mech.: Theor. Exp., P10002 (2006).

\bibitem{Wood1972} D. W. Wood and H. P. Griffiths, J. Phys. C \textbf{5},
L253 (1972).

\bibitem{Baxter1973} R. J. Baxter and F. Y. Wu, Phys. Rev. Lett. \textbf{31}, 1294 (1973).

\bibitem{Turban1982} L. Turban, J. Phys. C \textbf{15}, L65 (1982).

\bibitem{Turban1983} L. Turban and J. M. Debierre, J. Phys. A \textbf{16},
3571 (1983).

\bibitem{Potts1952} R. B. Potts, Proc. Camb. Phil. Soc. \textbf{48}, 106
(1952).

\bibitem{Wu1992} F. Y. Wu, Rev. Mod. Phys. \textbf{54}, 235 (1992)

\bibitem{Alcaraz1987} F. C. Alcaraz and M. N. Barber, J. Phys. A \textbf{20}, 179 (1987).

\bibitem{Baxter1982} R. J. Baxter, \textbf{Exactly Solved Models in Statistical
Mechanics}, Academic Press, New York, 1982.

\bibitem{Alcaraz1997} F. C. Alcaraz and J.C. Xavier, J. Phys. A \textbf{30},
L203 (1997).

\bibitem{Santos2001} M. Santos and W. Figueiredo, Phys. Rev. E \textbf{63},
042101 (2001).

\bibitem{Arashiro2003} E. Arashiro and J. R. Drugowich de Fel\'{\i}cio,
Phys. Rev. E \textbf{67}, 046123 (2003).

\bibitem{Chatelain2004} C. Chatelain, J. Stat. Mech.: Theor. Exp. P06006
(2004).

\bibitem{Igloi1983} F. Igl\'{o}i, D. V. Kapor, M. Skrinjar, and J.
S\'{o}lyom, J. Phys. A: Math. Gen. \textbf{16}, 4067 (1983).

\bibitem{Igloi1987} C. Vanderzande and F. Igl\'{o}i, J. Phys. A:
Math. Gen. \textbf{20}, 4539 (1987).

\bibitem{Jaster1999} A. Jaster, E. Manville, L. Schulke, and B. Zheng, J.
Phys. A: Math. Gen. \textbf{32}, 1395 (1999).

\bibitem{Schulke1995} L. Schulke and B. Zheng, Phys. Lett. A \textbf{204},
295 (1995).

\bibitem{Tome1998} T. Tom\'{e} and M. J. de Oliveira, Phys. Rev. E \textbf{58}, 4242 (1998).

\bibitem{Fernandes2005} H. A. Fernandes, J. R. Drugowich de Fel\'{\i}cio,
and A. A. Caparica, Phys. Rev. B. \textbf{72}, 054434 (2005).

\bibitem{Janssen1994} H. K. Janssen and K. Oerding, J. Phys. A: Math. Gen.
\textbf{27}, 715 (1994).

\bibitem{Dasilva2002} R. da Silva, N. A. Alves, and J. R. Drugowich de
Fel\'{\i}cio, Phys. Rev. E \textbf{66}, 026130 (2002).

\bibitem{Dasilva2004} R. da Silva and J. R. Drugowich de Fel\'{\i}cio, Phys.
Lett. A \textbf{333}, 277 (2004).

\bibitem{Simoes2001} C. S. Sim\~{o}es and J. R. Drugowich de Fel\'{\i}cio,
Mod. Phys. Lett. B \textbf{15}, 487 (2001).

\bibitem{Malakis2005} I. A. Hadjiagapiou, A. Malakis, and S. S. Martinos,
Physica A \textbf{356}, 563 (2005).

\bibitem{Tome2003} T. Tom\'{e}, J. Phys. A: Math. Gen. \textbf{36}, 6683
(2003).

\bibitem{Wegner1976} F. J. Wegner, \textbf{Phase Transitions and Critical Phenomena},
Vol. 6, edited by C. Domb and M.
S. Green, Academic Press, New York, 1976.

\bibitem{Kenna1986} R. Kenna, D. A. Johnston, and W. Janke, Phys.
Rev. Lett. \textbf{96}, 115701 (2006);

\bibitem{Reis1996} F. D. A. Aara\~{a}o Reis, S. L. A. de Queiroz, and Raimundo R.
dos Santos, Phys. Rev. B, \textbf{54}, R9616 (1996); \textbf{56},
6013 (1997).

\bibitem{Vanderzande} C. Vanderzande, J. Phys. A \textbf{20}, L549
(1987).
\end{thebibliography}
\end{document}